\documentclass[twocolumn]{aastex63}

\usepackage{lineno}
\usepackage[figuresright]{rotating}

\received{***, ****}
\revised{***, ****}
\accepted{***, ****}
\submitjournal{ApJL}

\begin{document}
\title{The Structure of Coronal Mass Ejections Recorded by the K-Coronagraph at Mauna Loa Solar Observatory}

\correspondingauthor{Hongqiang Song}
\email{hqsong@sdu.edu.cn}

\author{Hongqiang Song}
\affiliation{Shandong Provincial Key Laboratory of Optical Astronomy and Solar-Terrestrial Environment, and Institute of Space Sciences, Shandong University, Weihai, Shandong 264209, China}







\author{Leping Li}
\affiliation{National Astronomical Observatories, Chinese Academy of Sciences, Beijing, 100101, China}

\author{Zhenjun Zhou}
\affiliation{School of Atmospheric Sciences, Sun Yat-sen University, Zhuhai, Guangdong 519000, China}


\author{Lidong Xia}
\affiliation{Shandong Provincial Key Laboratory of Optical Astronomy and Solar-Terrestrial Environment, and Institute of Space Sciences, Shandong University, Weihai, Shandong 264209, China}


\author{Xin Cheng}
\affiliation{School of Astronomy and Space Science, Nanjing University, Nanjing, Jiangsu 210093, China}

\author{Yao Chen}
\affiliation{Shandong Provincial Key Laboratory of Optical Astronomy and Solar-Terrestrial Environment, and Institute of Space Sciences, Shandong University, Weihai, Shandong 264209, China}
\affiliation{Institute of Frontier and Interdisciplinary Science, Shandong University, Qingdao, Shandong 266237, China}







\begin{abstract}
Previous survey studies reported that coronal mass ejections (CMEs) can exhibit various structures in white-light coronagraphs, and $\sim$30\% of them have the typical three-part feature in the high corona (e.g., 2--6 $R_\odot$), which has been taken as the prototypical structure of CMEs. It is widely accepted that CMEs result from eruption of magnetic flux ropes (MFRs), and the three-part structure can be understood easily by means of the MFR eruption. It is interesting and significant to answer why only $\sim$30\% of CMEs have the three-part feature in previous studies. Here we conduct a synthesis of the CME structure in the field of view (FOV) of K-Coronagraph (1.05--3 $R_\odot$). In total, 369 CMEs are observed from 2013 September to 2022 November. After inspecting the CMEs one by one through joint observations of the AIA, K-Coronagraph and LASCO/C2, we find 71 events according to the criteria: 1) limb event; 2) normal CME, i.e., angular width $\geq$ 30$^{\circ}$; 3) K-Coronagraph caught the early eruption stage. All (or more than 90\% considering several ambiguous events) of the 71 CMEs exhibit the three-part feature in the FOV of K-Coronagraph, while only 30--40\% have the feature in the C2 FOV (2--6 $R_\odot$). For the first time, our studies show that 90--100\% and 30--40\% of normal CMEs possess the three-part structure in the low and high corona, respectively, which demonstrates that many CMEs can lose the three-part feature during their early evolutions, and strongly supports that most (if not all) CMEs have the MFR structures.
\end{abstract}



\section{Introduction}
On 1971 September 29, the first orbiting white-light coronagraph \citep{koomen75} was launched on Orbiting Solar Observatory Number 7 \citep{follett74}, and the coronagraph recorded the first images of a coronal mass ejection (CME) on 1971 December 14 \citep{tousey73}. From that time onwards, CMEs have been a subject of intense investigation in solar physics \citep{chenpengfei11,webb12}. Theoretically, CMEs originate from eruption of magnetic flux ropes (MFRs), which can form prior to \citep{lowbc01,patsourakos13,song15b,kliem21} and during \citep{song14a,ouyang15,wangwensi17,jiangchaowei21} solar eruptions.

In general, hot channels and coronal cavities are regarded as the proxies of MFRs in active regions \citep{zhangjie12,song15b} and quiet-Sun regions \citep{wangyimin08,chenyajie18}, respectively. The MFR can provide support to the prominence against gravity \citep[e.g.,][]{yanxiaoli16}. Therefore, CMEs are usually associated with eruptions of hot channels (high-temperature ejecta) \citep{zhangjie12,chengxin13a,song15c}, coronal cavities (middle-temperature ejecta) \citep{gibson06b,forland13,song22b}, and/or prominences (low-temperature ejecta) \citep{gopalswamy03,lileping12,song18a,zhouyuhao23,zhouzhenjun23} observationally.

The white-light coronagraph on the Solar Maximum Mission satellite recorded a CME with three-part structure on 1980 August 5, i.e., a bright core within a dark cavity surrounded by a bright loop front \citep{illing85} that included the low coronal observations of the CME from Mauna Loa MK3 coronameter. Since then, the three-part structure has become the prototypical structure of CMEs \citep[e.g.,][]{howardR06} though only $\sim$30\% of CMEs exhibit the three-part feature in the high corona, e.g., 2--6 $R_\odot$. In several decades, it is widely accepted that the bright front originates from the plasma pileup along the MFR boundary, the cavity represents the MFR, and the bright core corresponds to the prominence \citep[e.g.,][]{vourlidas13}.

However, recent studies pointed out that the traditional opinion is questionable, because some three-part CMEs are not associated with prominence eruptions at all \citep{howard17,song17b,song19b,song23a,wangbitao22}. Based on dual-viewpoint and seamless observations from the inner to outer corona, a new explanation on the three-part nature has been proposed, in which the bright frontal loop is formed due to the compression as the magnetic loops are successively pushed to stretch up by the underlying MFR \citep{lowbc01,chenpengfei09}, the core can correspond to the MFR and/or prominence, and the dark cavity between the CME front and the MFR is a low-density zone with sheared magnetic field \citep{song17b,song19a}. Recent observations clearly demonstrated that both hot channels \citep{song23a} and coronal cavities \citep{song22b} evolved into the bright core of three-part CMEs.

The new explanation also points out that CMEs can lose the three-part feature gradually when propagating outwards, because the dark cavity vanishes due to the MFR expansion and growth through magnetic reconnections, and/or because the bright core fades away due to the prominence expansion and drainage \citep{song23a}. This answers why only a portion of CMEs have the prototypical structure in previous survey studies \citep[e.g.,][]{vourlidas13}. As mentioned, CMEs result from MFR eruption, thus the new explanation predicts that all normal CMEs possess the three-part structure in the low corona, i.e., in the early eruption stage. Here the normal CMEs do not include the narrow ones with angular width less than 30$^{\circ}$ for two factors: 1) the narrow events might be jets, instead of CMEs with small angular width; 2) to observe the structure of narrow CMEs, coronagraphs with higher spatial resolution are necessary.

To examine whether all normal CMEs have the three-part structure in the early eruption stage, we conduct a survey study based on observations of the coronal solar magnetism observatory (COSMO) K-coronagraph (K-COR) from 2013 September to 2022 November. The paper is organized as follows. Section 2 introduces the related instruments and Methods. The observations and results are displayed in Section 3, which is followed by a summary and discussion in the final section.

\section{Instruments and Methods}
The Atmospheric Imaging Assembly (AIA) \citep{lemen12} on board the Solar Dynamics Observatory (SDO) \citep{pesnell12} takes images of the Sun through seven EUV channels. The AIA has a field of view (FOV) of 1.3 $R_\odot$, a spatial resolution of 1.2$\arcsec$ and a cadence of 12 s. Here we use the 131 \AA\ (Fe XXI, $\sim$10 MK) and 304 \AA\ (He II, $\sim$0.05 MK) to display the hot channel and prominence, respectively.

The K-COR is one of three proposed instruments in the COSMO facility suite \citep{tomczyk16} located at the Mauna Loa Solar Observatory (MLSO), and records the coronal polarization brightness (pB) in the passband of 7200--7500 \AA, which is formed by Thomson scattering of photospheric light from free electrons \citep{hayes01}. The FOV of K-COR is 1.05--3 $R_\odot$ with a pixel size of 5.5$\arcsec$ and a nominal cadence of 15 s. The Large Angle and Spectrometric Coronagraph \citep[LASCO;][]{brueckner95} on board the Solar and Heliospheric Observatory \citep[SOHO;][]{domingo95} comprises of three telescopes (C1, C2 and C3), each of which has an increasingly large FOV. Here the C2 (FOV: 2--6 $R_\odot$) is adopted to observe the CME structure in the outer corona.

The images of space-borne LASCO/C2 have better contrast than those of the ground-based K-COR. The normalized radially graded filter (NRGF) is employed to increase the K-COR contrast, which flattens the steep brightness gradient of the corona \citep{morgan06}. In this paper, we examine the CME structure through the NRGF data of K-COR that are available online\footnote{www2.hao.ucar.edu/mlso/mlso-home-page}, while for the C2 observations the original data are used.

\section{Observations and Results}
The K-COR data are available since 2013 September 30, while the coronagraph is closed temporarily due to the volcanic eruption of Mauna Loa on 2022 November 27. Therefore, our survey covers an interval from 2013 to 2022, during which 369 CMEs (excluding the possible events) are identified. On the whole, more CMEs are recorded around solar maximum though the K-COR does not work 24 hours continuously. The basic information for each event, such as the date, time (Universal Time, UT), and location (E--east, S--south, W--west, and N--north), is listed on the MLSO website. After inspecting the 369 CMEs combining observations of the AIA, K-COR and C2, we find 71 events according to the criteria: 1) limb event, which requires that the source region centered within 30$^{\circ}$ of the solar limb for the front-side events. For the far-side events, a limb event requires that the suspended prominence (or hot channel) prior to the eruption or the coronal disturbance (or post eruption arcade) during the eruption can be observed with the AIA; 2) normal CME, i.e., angular width $\geq$ 30$^{\circ}$ in the C2 FOV; 3) K-COR caught the early eruption stage.

We first scrutinize the 71 CMEs one by one through the NRGF images of K-COR, and find that all of them have the three-part feature in the low corona, irrespective of their appearance in the C2 images, agreeing with the prediction of the new explanation on the three-part structure of CMEs. However, visual identification of the three-part feature is not entirely objective as no quantitative criteria. There exist 6 events that do not have the clear three-part feature in the K-COR images (See Table 1), and several or all of them might be identified as the non-three-part CMEs. Therefore, we suggest that 90--100\% of the 71 CMEs possess the three-part structure in the low corona.

Table 1 lists the information of the 71 CMEs. The first column is the sequential number. Columns 2--4 give the date, time, and location of each event, which are from the MLSO website. The asterisks in Column 2 denote the 6 ambiguous events in the K-COR images mentioned above. Combining the observations of K-COR and AIA, we identify the type of source region, i.e., active region (AR) or quiet-Sun region (QS), and the ejecta, i.e., hot channel (HC) or prominence (P) for each event. The source type and ejecta are listed in Columns 5 and 6, respectively, where the ``?'' denotes that the source-region type or the ejecta are unsure, mainly because the events are located on the far side of the Sun. The subsequent four columns present CME information observed with LASCO, which are provided by the coordinated data analysis workshops (CDAW\footnote{https://cdaw.gsfc.nasa.gov}). Column 7 is the time of the CME's first appearance in the C2 FOV, and Columns 8--10 are the central position angle (PA), linear velocity (LV), as well as angular width (AW) correspondingly. The last column tells whether the CME exhibits the three-part feature in the C2 FOV, with Y/N denoting yes/no. Note that the asterisks in the last column indicate the ambiguous events in the C2 images.

\startlongtable
\centerwidetable
\begin{deluxetable*}{ccccccccccc}
	\tabletypesize{\small}
	\tablewidth{0pt}
	\tablecaption{The information of 71 limb CMEs in the K-COR and LASCO/C2 observations. Universal Time is used. \label{Table 1}}
	\tablehead{
		\colhead{No.} & \colhead{Date} & \colhead{K-Cor Time} &\colhead{Location} & \colhead{Source} & \colhead{Ejecta} & \colhead{First in C2} &\colhead{PA}& \colhead{LV}   & \colhead{AW} &\colhead{Three Part} \\
                      &    (yyyymmdd)  & (hhmm--hhmm)          &                   &                  &                  &  (hh:mm:ss)           &  ($\circ$) &  (km s$^{-1}$) &  ($\circ$)   &    in C2?
	}\startdata
	1  & 20140211 & 1845--1930 & W-SW & AR & P  & 19:24:05  & 248 & 613 & 271 & Y \\
	2  & 20140220 & 2224--2250 & W    & AR & P  & 23:12:11  & 282 & 198 & 45  & N \\
    3  & 20140429 & 1940--0046 & SW   & QS & P  & 20:57:25  & 229 & 232 & 71  & N \\
    4  & 20140524 & 2108--2200 & NE   & AR & P  & 22:00:05  & 66  & 377 & 180 & N \\
    5  & 20140528 & 1714--2118 & W-NW & QS & P  & 20:36:05  & 297 & 296 & 84  & Y \\
    6  & 20140614 & 1926--1940 & E-SE & AR & HC & 19:48:28  & 89  & 732 & 139 & N \\
    7  & 20140626 & 2114--0000 & NE   & AR & HC & 21:48:57  & 41  & 497 & 231 & N$^{*}$ \\
    8  & 20140630 & 1733--1848 & SW   & QS & P  & 18:36:05  & 261 & 262 & 72  & N \\
    9  & 20140923 & 2336--2349 & NE   & AR & HC & 00:48:05  & 68  & 311 & 52  & N \\
    10 & 20141014 & 1836--2002 & E-SE & AR & HC & 18:48:06  & 90  & 848 & 360 & Y \\
    11 & 20141105 & 1923--1950 & E-NE & AR & HC & 19:48:05  & 76  & 608 & 203 & N \\
    12 & 20141210$^{*}$ & 1749--1958 & SW   & AR & P  & 18:00:06  & 322 & 1086& 228 & N \\
    13 & 20141221 & 0048--0150 & E-NE & AR & P  & 01:25:53  & 60  & 283 & 116 & N \\
    14 & 20141221 & 0152--0219 & E-NE & AR & P  & 02:36:05  & 60  & 283 & 116 & Y \\
    15 & 20150208$^{*}$ & 2219--2250 & E    & AR & P  & 22:36:06  & 100 & 315 & 132 & N \\
    16 & 20150425 & 1803--1900 & W-NW & AR & P  & 18:48:05  & 304 & 493 & 50  & N \\
    17 & 20150501$^{*}$ & 2156--2228 & SW   & AR?& P  & 22:12:05  & 264 & 253 & 83  & N \\
    18 & 20150505 & 2209--2300 & NE   & AR & P  & 22:24:05  & 41  & 715 & 360 & N \\
    19 & 20150516 & 0101--0200 & NE   & QS & P  & 00:12:06  & 42  & 600 & 177 & Y \\
    20 & 20150525 & 2110--0045 & E    & QS & P  & 23:12:11  & 81  & 374 & 120 & N$^{*}$ \\
    21 & 20150702 & 1712--1904 & NE   & AR?& P  & 17:48:04  & 58  & 629 & 161 & Y \\
    22 & 20150801 & 1704--2200 & NE   & QS & P  & 17:36:04  & 65  & 472 & 67  & N \\
    23 & 20150923 & 1800--2000 & SE   & QS & P  & 18:36:04  & 106 & 565 & 99  & Y \\
    24 & 20151217 & 1922--2300 & NE   & QS & P  & 20:57:28  & 74  & 137 & 83  & N \\
    25 & 20160101 & 2256--2350 & SW   & AR & P  & 23:24:04  & 227 & 1730& 360 & Y \\
    26 & 20160115 & 1940--2158 & SW   & AR & HC?& 20:36:04  & 247 & 292 & 95  & N \\
    27 & 20160208 & 2220--0032 & NE   & QS & P  & 22:00:06  & 18  & 311 & 164 & Y \\
    28 & 20160209 & 1758--2050 & E    & QS & P  & 19:23:30  & 72  & 358 & 76  & N \\
    29 & 20160611 & 2206--2238 & E-NE & AR & P  & 23:24:05  & 68  & 95  & 32  & N \\
    30 & 20160808 & 2022--2100 & W    & AR?& HC?& 20:48:06  & 260 & 674 & 84  & Y \\
    31 & 20160808 & 1900--0130 & W-NW & QS?& P? & 01:25:43  & 311 & 128 & 66  & N$^{*}$ \\
    32 & 20160929 & 1718--2050 & SW   & AR & HC & 20:12:05  & 254 & 447 & 125 & Y \\
    33 & 20170128$^{*}$ & 1928--0216 & W    & AR & HC & 21:48:05  & 280 & 562 & 53  & N \\
    34 & 20170327 & 1732--2010 & E    & AR & HC & 18:12:05  & 89  & 230 & 46  & N \\
    35 & 20170402 & 1844--2020 & W-NW & AR & HC & 19:24:05  & 290 & 500 & 88  & N \\
    36 & 20170713 & 2008--2026 & W    & AR & P  & 20:36:05  & 260 & 290 & 61  & N \\
    37 & 20170720 & 1700--1902 & W    & AR?& P? & 18:12:05  & 265 & 590 & 95  & Y \\
    38 & 20170820 & 1934--2104 & E    & AR & HC & 20:24:05  & 88  & 207 & 43  & N \\
    39 & 20170912 & 1858--1941 & W-SW & AR?& HC?& 19:12:05  & 271 & 476 & 113 & Y \\
    40 & 20171020 & 2328--0014 & SE   & AR & P  & 00:00:05  & 98  & 331 & 109 & N \\
    41 & 20190422$^{*}$ & 0254--0422 & W-NW & QS?& P  & 03:24:05  & 269 & 422 & 55  & N \\
    42 & 20201101 & 1900--2020 & SW   & AR & HC & 19:48:05  & 266 & 289 & 36  & N \\
    43 & 20201126 & 2020--2105 & NE   & AR & HC & 21:12:10  & 99  & 572 & 92  & Y \\
    44 & 20210429 & 1701--2146 & NE   & QS & P  & 20:01:34  & 75  & 189 & 129 & N$^{*}$ \\
    45 & 20210507 & 1852--2005 & E    & AR & HC & 19:24:05  & 76  & 754 & 114 & N \\
    46 & 20210610 & 1746--1904 & E-NE & AR?& P  & 18:24:05  & 83  & 833 & 133 & Y \\
    47 & 20210625 & 2017--0145 & SW   & QS & P  & 00:48:05  & 234 & 101 & 30  & N \\
    48 & 20210626 & 2132--0210 & W-NW & AR & P  & 03:48:05  & 247 & 223 & 115 & N \\
    49 & 20210715 & 2110--2344 & SE   & QS & P  & 21:36:05  & 166 & 1476& 360 & Y \\
    50 & 20210719$^{*}$ & 2022--2122 & E    & AR & HC & 20:57:05  & 69  & 401 & 133 & N$^{*}$ \\
    51 & 20210829 & 1956--2040 & SW   & AR & P  & 20:24:05  & 259 & 1060& 87  & N \\
    52 & 20211009 & 2030--0125 & W-NW & AR & HC & 22:36:05  & 275 & 433 & 110 & Y \\
    53 & 20211010 & 2235--0150 & NW   & AR & P  & 23:24:05  & 293 & 299 & 71  & N \\
    54 & 20211102 & 2130--2353 & SE   & AR & HC & 22:24:05  & 113 & 474 & 98  & N \\
    55 & 20211103 & 2050--2216 & W-SW & AR & P  & 21:36:05  & 260 & 510 & 360 & N \\
    56 & 20220131 & 2328--0001 & SW   & QS & P  & 00:12:05  & 245 & 469 & 52  & N \\
    57 & 20220201 & 2300--0213 & SW   & QS & P  & 01:25:48  & 243 & 467 & 43  & N$^{*}$ \\
    58 & 20220419 & 2043--2204 & W-SW & AR & HC & 21:24:05  & 227 & 247 & 53  & N \\
    59 & 20220425 & 1723--1854 & SE   & AR & P  & 18:00:05  & 85  & 319 & 120 & N \\
    60 & 20220425 & 1836--2131 & SE   & AR & P  & 20:24:05  & 90  & 498 & 125 & Y \\
    61 & 20220508 & 2124--2250 & SE   & AR & HC & 22:24:05  & 94  & 602 & 175 & Y \\
    62 & 20220511 & 1820--1950 & W    & AR & P  & 18:36:05  & 237 & 1100& 194 & N \\
    63 & 20220514 & 1653--2106 & SE   & QS & P  & 18:24:05  & 115 & 843 & 52  & N \\
    64 & 20220524 & 2219--2324 & NE   & AR & P  & 23:12:11  & 71  & 569 & 211 & Y \\
    65 & 20220710 & 1658--1931 & SW   & QS & P  & 17:48:05  & 198 & 1241& 43  & N \\
    66 & 20220731 & 2236--0010 & E    & AR & P  & 23:12:10  & 82  & 1122& 192 & N \\
    67 & 20220830 & 1744--1927 & S-SW & AR & HC & 18:12:05  & 268 & 1247& 360 & Y \\
    68 & 20220924 & 1957--2339 & SE   & QS & P  & 20:24:05  & 132 & 337 & 103 & N$^{*}$ \\
    69 & 20220928 & 1718--1818 & E    & AR & P  & 17:36:05  & 87  & 256 & 115 & N \\
    70 & 20221026 & 1900--2231 & SW   & QS & P  & 21:12:09  & 207 & 506 & 167 & N \\
    71 & 20221125 & 2128--2356 & NW   & QS & P  & 22:24:05  & 312 & 620 & 106 & N \\
	\enddata
		
\end{deluxetable*}
\vspace{-0.5cm}
Table 1 shows that 49 and 22 CMEs originate from active regions (AR) and quiet-Sun regions (QS), respectively. For the ejecta type, there are 49 events associated with a prominence (P) eruption, and the rest are correlated with a hot-channel (HC) eruption. In the LASCO FOV, the linear velocities range from 95 to 1730 km s$^{-1}$, and their angular widths, from 30 to 360$^{\circ}$. After scrutinizing the 71 CMEs one by one through the C2 images, we find that only 21 events ($\sim$30\%) possess the three-part structure, agreeing with previous statistical results based on C2 observations \citep[e.g.,][]{vourlidas13}. For the 50 non-three-part CMEs in the C2 images, 7 ones are ambiguous and might be identified as the three-part events, which are denoted with asterisks in the last column as mentioned. Therefore, we suggest that 30--40\% of the 71 CMEs possess the three-part structure in the high corona. To demonstrate that a hot channel or prominence can lead to a three-part CME in its early eruption stage, and the three-part feature can sustain or disappear in the outer corona, we select four representative events and display them in Figures 1--4 sequentially.

Figure 1 displays the event occurring on 2014 October 14 (Event 10 in Table 1), which resulted from a hot-channel eruption in an active region located at the SE limb. Panel (a) shows the hot channel with the AIA 131 \AA\ observation at 18:45:32 UT. Panel (b) presents the K-COR observation at 18:58:51 UT, and the three-part CME can be identified. The bright core is very obvious, and the bright front is delineated with the red-dashed line as we can not discern the front clearly in the static image. The bright front and three-part structure can be distinguished clearly through the animation accompanying with Figure 1. The C2 image at 20:00:05 UT is presented in Panel (c), and we can see that the CME keeps the typical three-part feature there. Note that the white circles in both Panels (b) and (c) denote the solar limb.

Figure 2 presents the CME occurring on 2016 January 1 (Event 25 in Table 1). This event is associated with a prominence eruption that can be observed in the AIA 304 \AA\ image as shown in Panel (a). Panel (b) displays the K-COR observation at 23:23:42 UT, in which the bright front and core are delineated with the red- and yellow-dashed lines, respectively, to display the three-part structure clearly. Please see the accompanied animation to examine the three-part feature continuously. This CME also exhibits the three-part feature in the C2 FOV as shown in Panel (c), where the red-dashed line denotes the leading front.

The above two CMEs exhibit the three-part structure in both K-COR and C2 images. Next we'll show two events that do not have the three-part feature in the C2 FOV. Figure 3 displays a CME induced by the hot-channel eruption as revealed with the AIA 131 \AA\ observation in Panel (a). This event occurred on 2021 May 7 (Event 45 in Table 1) and a previous study  \citep{wangbitao22} has demonstrated that the hot channel evolved into the bright core in the K-COR image as shown in Panel (b). The three-part feature is distinguishable directly in the static image, thus no dashed lines are guide for the eye in this panel. Panel (c) shows the observation of C2 at 19:48:05 UT with the red-dashed line delineating the CME front. The CME lost its three-part feature due to the MFR expansion and growth as mentioned \citep{song19a,song23a}.

Figure 4 shows the CME occurring on 2021 October 10 (Event 53 in Table 1), which is associated with the prominence eruption from an active region located at the NW limb. Panel (a) illustrates the erupting prominence with the AIA 304 \AA\ image at 22:51:29 UT. The K-COR observations present a typical three-part CME as shown in Panel (b), where the red-dashed line depicts the CME front. Please see the accompanied animation to view the three-part feature in the K-COR images. The animation also demonstrates that the prominence did not erupt outward eventually. This leads to a non-three-part CME in the C2 image \citep{song23a} as presented in Panel (c), where the red-dashed line delineates the CME front.

\section{Summary and Discussion}
To verify the new explanation on the three-part structure of CMEs, which predicts that all normal CMEs should exhibit the three-part feature in their early eruption stage, we conducted a survey study on the CME structure with the observations of K-COR (FOV: 1.05--3 $R_\odot$) at MLSO. In total, 369 CMEs (excluding the possible events) are identified from 2013 September to 2022 November. Combining the observations of AIA, K-COR and LASCO/C2, we inspected the events one by one manually, and found 71 events according to the criteria: 1) limb event; 2) normal CME with angular width $\geq$ 30$^{\circ}$ in the C2 FOV (2--6 $R_\odot$); 3) K-COR caught the early eruption stage. The results showed that 90--100\% of the 71 events exhibit the three-part structure in the K-COR observations, basically agreeing with the prediction of the new explanation, and 30--40\% of the events have the three-part appearance in the C2 observations, consistent with previous survey studies \citep[e.g.,][]{vourlidas13,song23a}. These suggest that CMEs can lose the three-part feature during their propagation outwards, and further support the new explanation on the nature of the three-part structure of CMEs\citep{song22b}.

As mentioned, theoretical studies demonstrate that CMEs result from MFR eruption, and no physical mechanism can produce large-scale CMEs without involving an MFR. However, do all CMEs have an MFR structure near the Sun observationally? Our current survey study intends to answer ``Yes" to this question, as 90--100\% of normal CMEs exhibit the three-part structure in their early eruption stage. We think that the several ambiguous events could exhibit the three-part feature if the observations were clearer. For the narrow CMEs (angular width $\textless$ 30$^{\circ}$), we speculate that they can also exhibit the three-part feature when the spatial resolution of coronagraphs is high enough.

The CME and MFR are called ICME (interplanetary CME) and magnetic cloud \citep{burlaga81}, respectively, after they leave the corona. If the MFR structures are not destroyed during their propagation, all ICMEs should possess the magnetic cloud features near 1 au, such as enhanced magnetic-field intensity, large and smooth rotation of the magnetic-field direction, and low proton temperature or low plasma $\beta$ \citep{zurbuchen06,wucc11,song20b}. However, only about one third of ICMEs have the magnetic cloud features near the Earth \citep{chiyutian16}. From the statistical point of view, this might be a result of glancing cuts between the spacecraft and ICME, as ICMEs with magnetic cloud features have narrower sheath region compared to the non-cloud ICMEs \citep{song20a}. The analyses on the morphological structure of CMEs near the Sun and the geometric character of ICMEs near 1 au support that most (if not all) CMEs have the MFR structures.


\acknowledgments We thank the anonymous referee for the comments and suggestions that helped to improve the original manuscript. We are grateful to Profs. Jie Zhang (GMU), Pengfei Chen (NJU), Yuandeng Shen (YNAO) and Mr. Zihao Yang (PKU) for their helpful discussions. This work is supported by the National Key R\&D Program of China 2022YFF0503003 (2022YFF0503000), the NSFC grants U2031109, 11790303 (11790300), and 12073042. H.Q.S is also supported by the CAS grants XDA-17040507. The authors acknowledge the use of data from the SDO, MLSO, and SOHO, as well as the usage of the CDAW CME catalog generated by NASA and The Catholic University of America and the MLSO activity tables created by the MLSO team.





\clearpage


\begin{figure*}[htb!]
\epsscale{1.1} \plotone{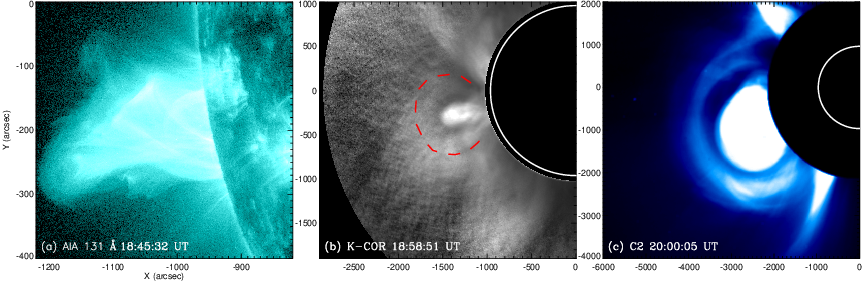} \caption{Observations of the CME occurred on 2014 October 14. (a) AIA 131 \AA\ image, showing the erupting hot channel at 18:45:32 UT. (b) NRGF image of K-COR at 18:58:51 UT, displaying the three-part structure in the low corona. The animation accompanying with this panel demonstrates the complete eruption process from 18:30:32 to 20:01:34 UT. The duration of the animation is 4 s. (c) LASCO/C2 observation at 20:00:05 UT, displaying the CME structure in the high corona. The white circles in both Panels (b) and (c) denote the solar limb. The red-dashed line delineates the CME front. \label{Figure 1}}
\end{figure*}

\begin{figure*}[htb!]
\epsscale{1.1} \plotone{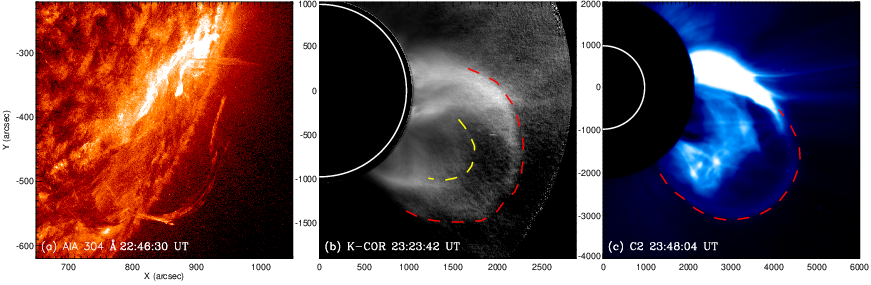} \caption{Observations of the CME occurred on 2016 January 1. (a) AIA 304 \AA\ image, showing the erupting prominence at 22:46:30 UT. (b) NRGF image of K-COR at 23:23:42 UT, displaying the three-part structure in the low corona. The animation accompanying with this panel demonstrates the complete eruption process from 22:49:34 to 23:51:31 UT. The duration of the animation is 2 s. (c) LASCO/C2 observation at 23:48:04 UT, displaying the CME structure in the high corona. The white circles in both Panels (b) and (c) denote the solar limb. The red-dashed line delineates the CME front, and the yellow-dashed line, the CME core. \label{Figure 2}}
\end{figure*}

\begin{figure*}[htb!]
\epsscale{1.1} \plotone{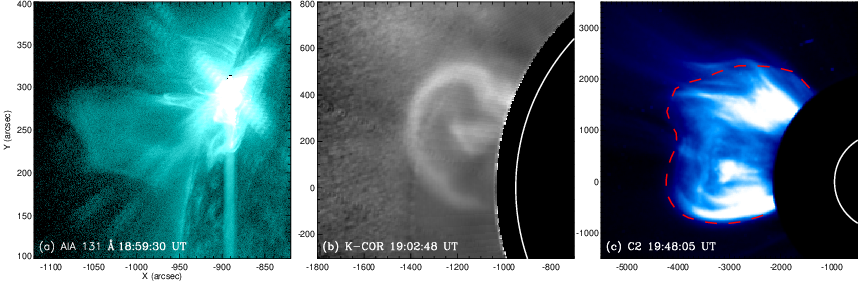} \caption{Observations of the CME occurred on 2021 May 7. (a) AIA 131 \AA\ image, showing the erupting hot channel at 18:59:30 UT. (b) NRGF image of K-COR at 19:02:48 UT, displaying the three-part structure in the low corona. The animation accompanying with this panel demonstrates the complete eruption process from 18:50:40 to 19:51:21 UT. The duration of the animation is 3 s. (c) LASCO/C2 observation at 19:48:05 UT, displaying the CME structure in the high corona.The white circles in both Panels (b) and (c) denote the solar limb. The red-dashed line delineates the CME front.  \label{Figure 3}}
\end{figure*}

\begin{figure*}[htb!]
\epsscale{1.1} \plotone{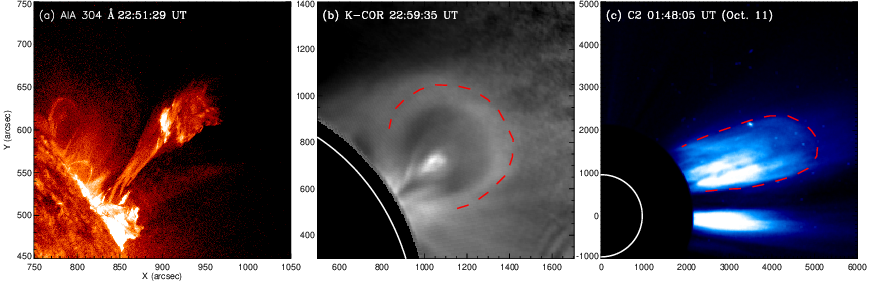} \caption{Observations of the CME occurred on 2021 October 10. (a) AIA 304 \AA\ image, showing the erupting prominence at 22:51:29 UT. (b) NRGF image of K-COR at 22:59:35 UT, displaying the three-part structure in the low corona. The animation accompanying with this panel demonstrates the complete eruption process from 22:28:08 to 23:50:09 UT. The duration of the animation is 4 s. (c) LASCO/C2 observation at 01:48:05 UT on October 11, displaying the CME structure in the high corona. The white circles in both Panels (b) and (c) denote the solar limb. The red-dashed line delineates the CME front. \label{Figure 4}}
\end{figure*}
\end{document}